\documentclass[a4paper,11pt]{article}
\pdfoutput=1 

\usepackage{jcappub} 

\newcommand{\beq}{\begin{equation}}

\newcommand{\degree}{\ensuremath{^\circ}}

\title{Galactic magnetic deflections and Centaurus A as a UHECR source}

\author[a,1]{Glennys R. Farrar,\note{Corresponding author.}}
\author[a]{Ronnie Jansson,}
\author[b]{Ilana J. Feain,}
\author[c]{B. M. Gaensler}

\affiliation[a]{Center for Cosmology and Particle Physics,
and Department of Physics,\\ New York University, NY, NY 10003, USA}
\affiliation[b]{CSIRO Astronomy \& Space Science, P.O. Box 76, Epping NSW 1710, Australia}
\affiliation[c]{Sydney Institute for Astronomy (SIfA), School of Physics, The University of Sydney, NSW 2006, Australia}

\emailAdd{gf25@nyu.edu}
\emailAdd{rj486@nyu.edu}
\emailAdd{Ilana.Feain@csiro.au}
\emailAdd{bryan.gaensler@sydney.edu.au}

\abstract{We evaluate the validity of leading models of the Galactic magnetic field for predicting UHECR deflections from Cen A.  The Jansson-Farrar 2012 GMF model (JF12), which includes striated and random components as well as an out-of-plane contribution to the regular field not considered in other models, gives by far the best fit globally to all-sky data including the WMAP7 22 GHz synchrotron emission maps for $Q,\,U \,\& \, I$ and $\approx 40,000$ extragalactic Rotation Measures (RMs).  Here we test the models specifically in the Cen A region, using 160 well-measured RMs and the Polarized Intensity from WMAP, nearby but outside the Cen A radio lobes.  The JF12 model predictions are in excellent agreement with the observations, justifying confidence in its predictions for deflections of UHECRs from Cen A.  We find that up to six of the 69 Auger events above 55 EeV are consistent with originating in Cen A and being deflected $\leq 18^{\circ}$;  in this case three are protons and three have $Z=2-4$.  Others of the 13 events within $18^{\circ}$ must have another origin.  In order for a random extragalactic magnetic field between Cen A and the Milky Way to appreciably alter these conclusions, its strength would have to be $\gtrsim 80$ nG -- far larger than normally imagined.}

\begin{document}
\maketitle7
\flushbottom
\section{Introduction}
\label{intro}

\begin{figure}
\centering
\includegraphics[width=.95\linewidth]{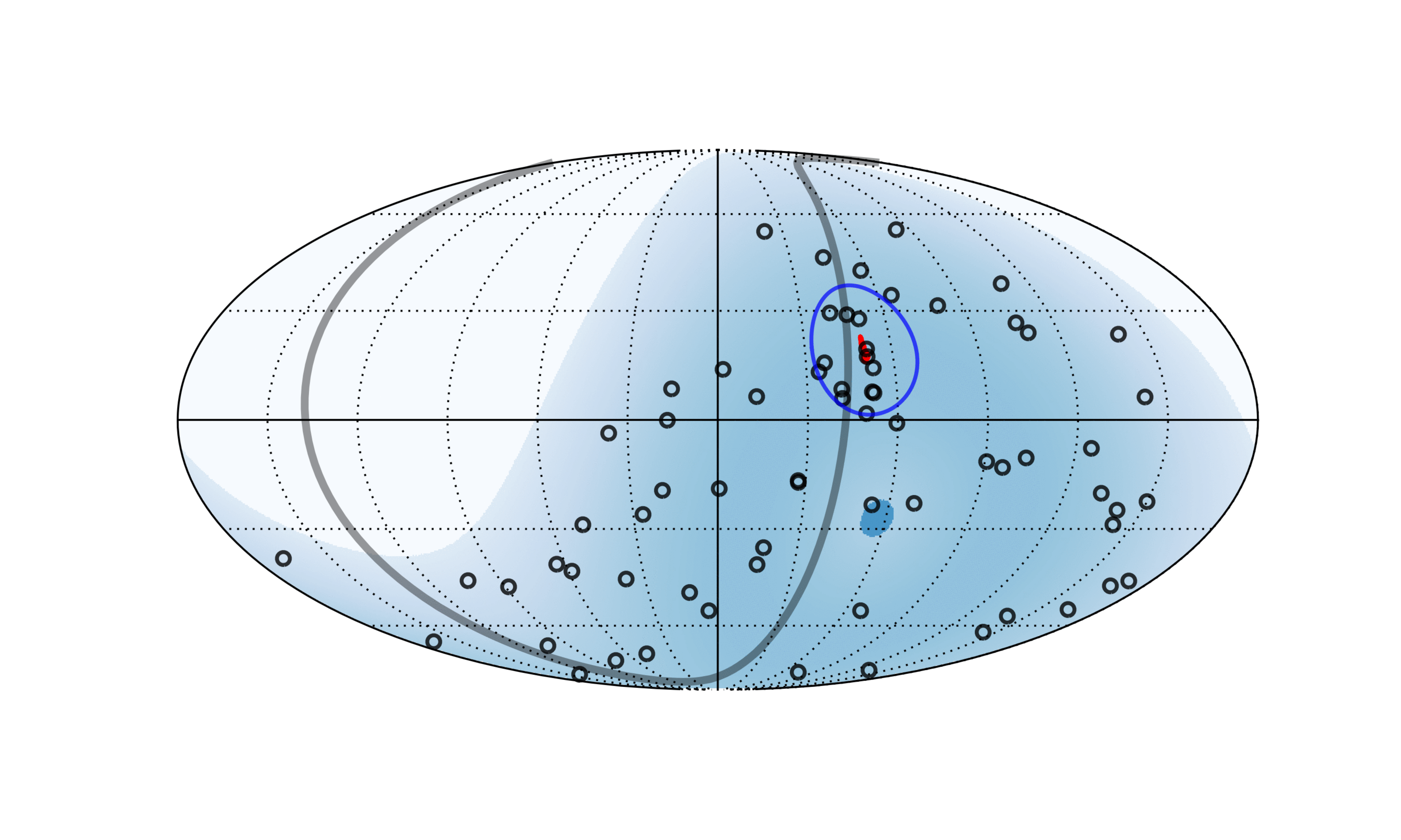}
\caption{The 69 published Auger events above 55 EeV\cite{Auger:2010} in Galactic coordinates, with Galactic longitude zero at the center and increasing to the left. The giant radio lobes of Cen A are marked in red and the super-Galactic plane is shown as a grey line. The blue contour is an $18^\circ$ circle around Cen A.  The Auger exposure is shown in light blue;  it is largest at the South Celestial Pole, about 50$^\circ$ south of Cen A. }\label{auger_events}
\end{figure}

The astrophysical origin of the highest energy cosmic rays remains one of the major open questions in astrophysics.  Cosmic rays with energies above about $55\times 10^{18}$ eV (55 EeV) -- commonly called UHECRs -- lose energy through interactions with the Cosmic Microwave Background radiation, implying that their sources must be relatively nearby (closer than $\sim 200$ Mpc\cite{Greisen:1966,Zatsepin:1966}).  However, very few classes of astrophysical systems seem capable of accelerating particles to such high energies and there are few candidate sources within the required distance.  The closest plausible UHECR source is the nearby radio galaxy Centaurus A\cite{Cavallo:1978,Romero:1996,Farrar:2000}.  Tantalizingly, the highest energy Pierre Auger Observatory events show a distinct excess within $18^\circ$ of Cen A: 13  events observed, while 3.2 events would be expected for 69 events coming from an isotropic distribution\cite{Auger:2010}.  This excess has re-ignited interest in the possibility that Cen A is the source of most of these events\cite{Wibig:2007,Gorbunov:2008,Hardcastle:2009,Kachelriess:2009,Rachen:2008, Fargion:2008,Fargion:2009}.  However, Cen A is located in the ``Supergalactic Plane'' -- a nearby overdensity of galaxies -- so the UHECR excess could be produced by multiple sources in the general region, or possibly simply be due to some focussing effect of the GMF or EGMF.  Fig. \ref{auger_events} shows the arrival directions of the 69 published events, with Cen A and the supergalactic plane indicated as well.  

\begin{figure}
\centering
\includegraphics[width=1\linewidth]{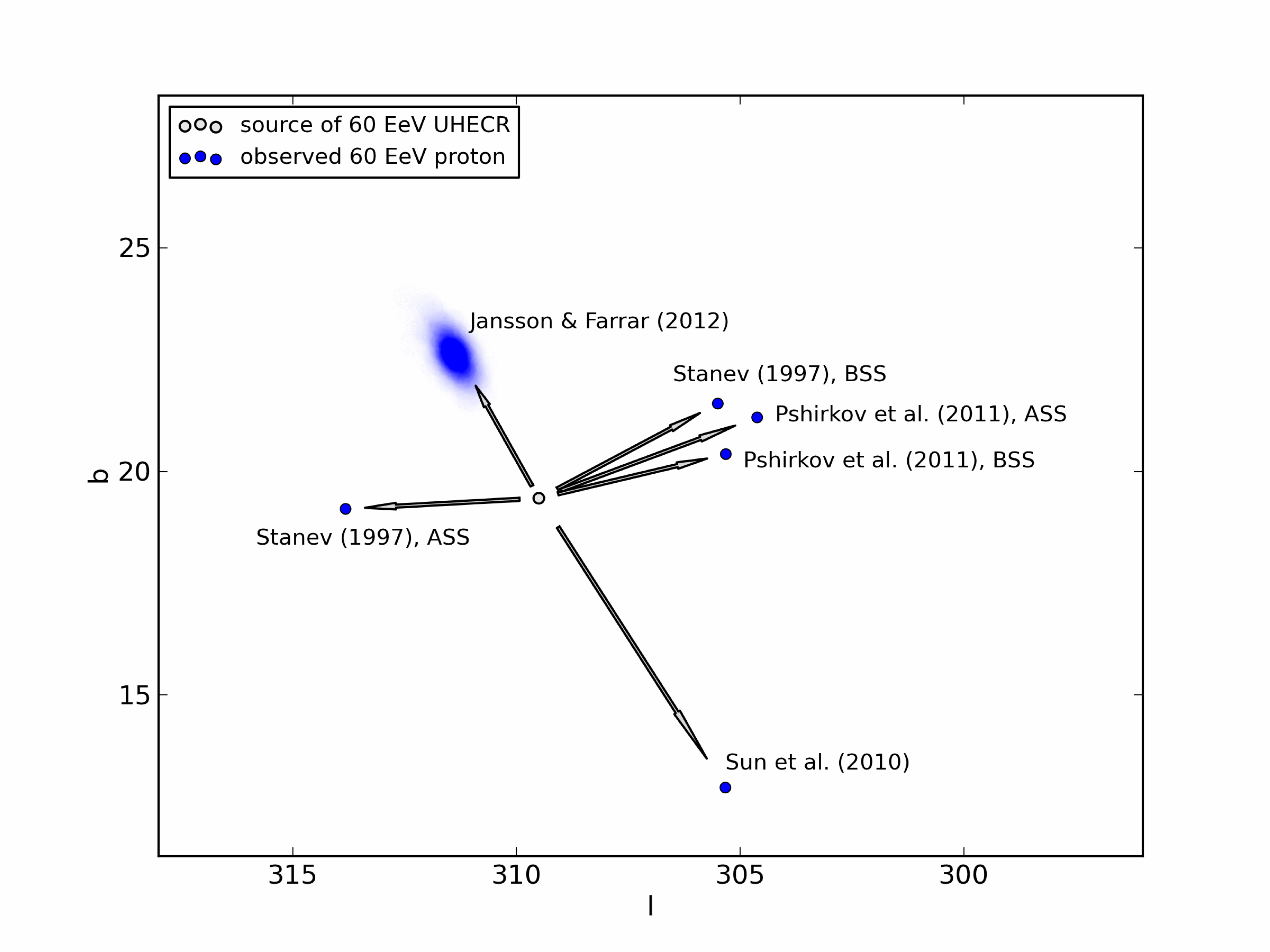}
\caption{The predicted locus of arrival directions for a 60 EeV proton emitted from the nucleus of Cen A (white circle), for the JF12 GMF and five other popular large-scale Galactic magnetic field models:  the ASS/BSS  models by Stanev \cite{Stanev:1997}, the best-fit model of Sun et al. \cite{Sun:2008,Sun:2010} with a 2 $\mu$G halo field, and the ASS/BSS models of Pshirkov et al. \cite{pshirkov+11}. The $2-\sigma$ uncertainty region of the predicted arrival direction due to the uncertainty in the JF12 parameter values is indicated by the shaded region; no such uncertainty analysis exists for the other models.  JF12 provides a model of the random field, but for purposes of comparing to the other models which do not provide a model of the random field, only deflections due to the coherent field are shown.}\label{stanev}
\end{figure}

Since UHECRs are charged protons or nuclei\cite{AugerICRCcorrelations:2009}, their arrival directions do not point exactly toward the source of their emission.  Magnetic deflection is proportional to the UHECR's electric charge, which is undetermined.  Some evidence, e.g., correlations with extragalactic objects such as AGNs \cite{AugerAGN:2007,Auger:2010}, favors UHECRs being mostly protons, while other evidence suggests a mixed or heavy composition \cite{AugerICRCcomposition:2009}.  Therefore, use of a reliable model of the Galactic magnetic field including both the large scale and random components, is necessary to obtain trustworthy predictions for UHECR source locations.  This is not a pedantic matter, as demonstrated in Fig. \ref{stanev} which shows the striking variation in predicted arrival directions using six currently-used models for the coherent GMF, for a 60 EeV proton produced in the core of Cen A.  

Of the six GMF models used in Fig. \ref{stanev}, the new model of Jansson and Farrar \cite{jf12,jf12rand} (JF12 below) gives by far the best global fit to the RM and polarized synchrotron data \cite{jf12}, with a $\chi_{\rm dof}^2$ per degree of freedom of 1.096 for the 6605 observables (pixels of RM, $Q$ and $U$).   The next best model is that of Sun et al.\cite{Sun:2008}, whose parameters were updated in \cite{Sun:2010} (SR10).  This is the most comprehensive attempt prior to JF12 to model the coherent GMF using constraints from both RM and synchrotron emission data.  It does not allow for as general a functional form as JF12, and in particular does not include the out-of-plane field or possible striated random fields.  With the parameters given in ref. \cite{Sun:2010}, its $\chi_{\rm dof}^2$ for the 6605 observables is 1.672.  To test if the functional form adopted in SR10 is as good as JF12, we used the SR10 form and re-optimized its parameters to fit the ensemble of the data using the JF12 MCMC;  the resulting fit has  $\chi_{\rm dof}^2 = 1.325$, indicating that the out-of-plane and striated field components of JF12 are signficant improvements in the model.  Although more recent, the models proposed by Pshirkov et al.\cite{pshirkov+11} (P+11) have a less general form and are less well constrained, as these authors used only RMs and not synchrotron emission.  They are based on the SR10 and Prouza-Smida\cite{Prouza2003} (PS03) models.  Being unable to disambiguate the large scale geometry, P+11 offers benchmark BSS and ASS versions.  When fitting the complete set of 6605 observables, these give $\chi_{\rm dof}^2 = 2.663, \, 4.971$, respectively; with our re-optimization of their parameter values these become $\chi_{\rm dof}^2 =  1.452, \, 1.591$.  We did not measure the $\chi_{\rm dof}^2$ of PS03 because the P+11 models are a generalization of it, and those give poor fits. The ASS and BSS models of Stanev\cite{Stanev:1997} are classics, developed to illustrate the impact of different field geometries more than to provide a detailed model for the field; they fare even worse in a global fit.   Clearly, studies of the deflection of UHECRs such as refs \cite{Takami:2009, Vorobiov:2009} using these old models, cannot be trusted to reliably predict CR deflections in the direction of Cen A. 

In Sec. \ref{gmf} we take advantage of the exceptional RM coverage in the region surrounding Cen A from \cite{Feain:2009}, to test the various GMF models in the region relevant for predicting deflections of UHECRs from Cen A.  We find that JF12 accurately predicts the mean Faraday rotation measure and polarized and total synchrotron intensity in the particular direction of Cen A, while other models perform less-well to very-poorly.  

Finally, having confirmed the validity of the JF12 model for Cen A deflections, we use JF12 in Sec. \ref{deflections} to determine the deflections of UHECRs through the GMF as a function of their energy and charge.  We find that three events within $18^{\circ}$ of Cen A could be protons coming from Cen A and three others can be attributed to Cen A for more general charge assignments.   Thus we find that the distribution of the arrival directions of the excess of events is not compatible with their dominant source being either the Active Galaxy or the extended radio lobes of Cen A, unless high Z nuclei can ``wrap back" to the Cen A region -- winding up arriving from that direction after deflections greater than $2 \pi$.  Of course, in that case, an association with Cen A would be essentially accidental.

\section{Centaurus A}

Centaurus A (NGC 5128) is the nearest active galaxy, and a Fanaroff-Riley Class I (FR-I) radio galaxy  (see \cite{Israel:1998} for a review), at a distance of 3.8 Mpc \cite{Harris:2009}. The massive elliptical host galaxy has Galactic coordinates $(l,\,b)=(309.5^\circ,\,19.4^\circ)$. Thanks to its proximity and size, its enormous radio lobes combine into the largest extragalactic object on the sky, with an angular size of 9\degree$\times$5\degree, corresponding to a physical size of 500$\times$250 kpc. About 5 kpc from the central galaxy, jets from the accretion disk surrounding the central supermassive black hole expand into plumes as they plow into the ambient intergalactic medium. These plumes are called the inner radio lobes. Some material goes farther, creating the northern middle lobe, which extends to 30 kpc and lacks a southern counterpart. The giant outer radio lobes extend 250 kpc in projection both in the north and the south; their outline is shown in Fig. \ref{cenA_data}. The 3D orientation of the lobes is not well-known.

Cen A was first considered as a possible source of UHECRs by Cavallo \cite{Cavallo:1978}. The possibility that Cen A could in fact be the source of most cosmic rays, if turbulent extragalactic magnetic fields on the Cen A side of the Milky Way are near the maximum allowed value, was first proposed by Farrar and Piran \cite{Farrar:2000}.   Some of the more recent works include: \cite{Wibig:2007}, which proposes that Cen A is one of three sources that, combined, are responsible for all observed UHECRs;  \cite{Gorbunov:2008} which analyzes the significance of correlation between UHECRs and Cen A; \cite{Hardcastle:2009}  which investigates the plausibility of the giant radio lobes of Cen A being acceleration sites of UHECRs; \cite{Kachelriess:2009} which considers the possibility that the radio jet at the core of Cen A is an accelerator; \cite{Rachen:2008, Rieger:2009} which considers various mechanisms that could accelerate UHECRs in radio galaxies such as Cen A; and \cite{Fargion:2008,Fargion:2009} which argue that Cen A is the source of the $\approx\!10$ events in the region of the sky surrounding Cen A. Ref. \cite{Moskalenko:2009} notes that Cen A could be associated with at least 4 out of the -- at the time -- 27 published Pierre Auger UHECRs above 57 EeV due to its large radio extent, but does not consider the deflection by any particular Galactic magnetic field model.

\section{Magnetic field model and predictions}\label{gmf}

As seen in Fig. \ref{stanev}, different GMF models yield highly disparate predictions for UHECR deflections. Hence, the validity of any conclusions about UHECR deflections hinges on the reliability of its GMF model.   The JF12 model of refs. \cite{jf12,jf12rand} should be substantially more realistic than earlier models.  It has a more general form and it is constrained by both RMs and synchrotron emission, which together probe both the line-of-sight and transverse components of the field.  The model includes a thin disk component, an extended halo component, and an out-of-plane component as suggested by observations of external galaxies; random and striated random fields are also included in the analysis. We refer the reader to \cite{jf12} for details of the JF12 large scale GMF model, and to \cite{jf12rand} for a description of the associated random field model.  The model is constrained by the WMAP7 Galactic synchrotron emission map \cite{Gold:2011} and more than forty thousand extragalactic rotation measures, and as noted in Sec. \ref{intro} it reproduces the global RM and polarized and unpolarized synchrotron data well.  

However, the Galactic magnetic field is very complicated and even the JF12 global model with 34 parameters describing the coherent and random fields cannot be expected to provide a highly accurate model of the magnetic field along every line-of-sight.   Therefore, before proceeding to UHECR deflections, we first determine the constraining observables along lines-of-sight relevant for UHECRs propagating from Cen A and compare them to the predictions of the JF12 model.  

\begin{figure}
\centering
\includegraphics[width=1\linewidth]{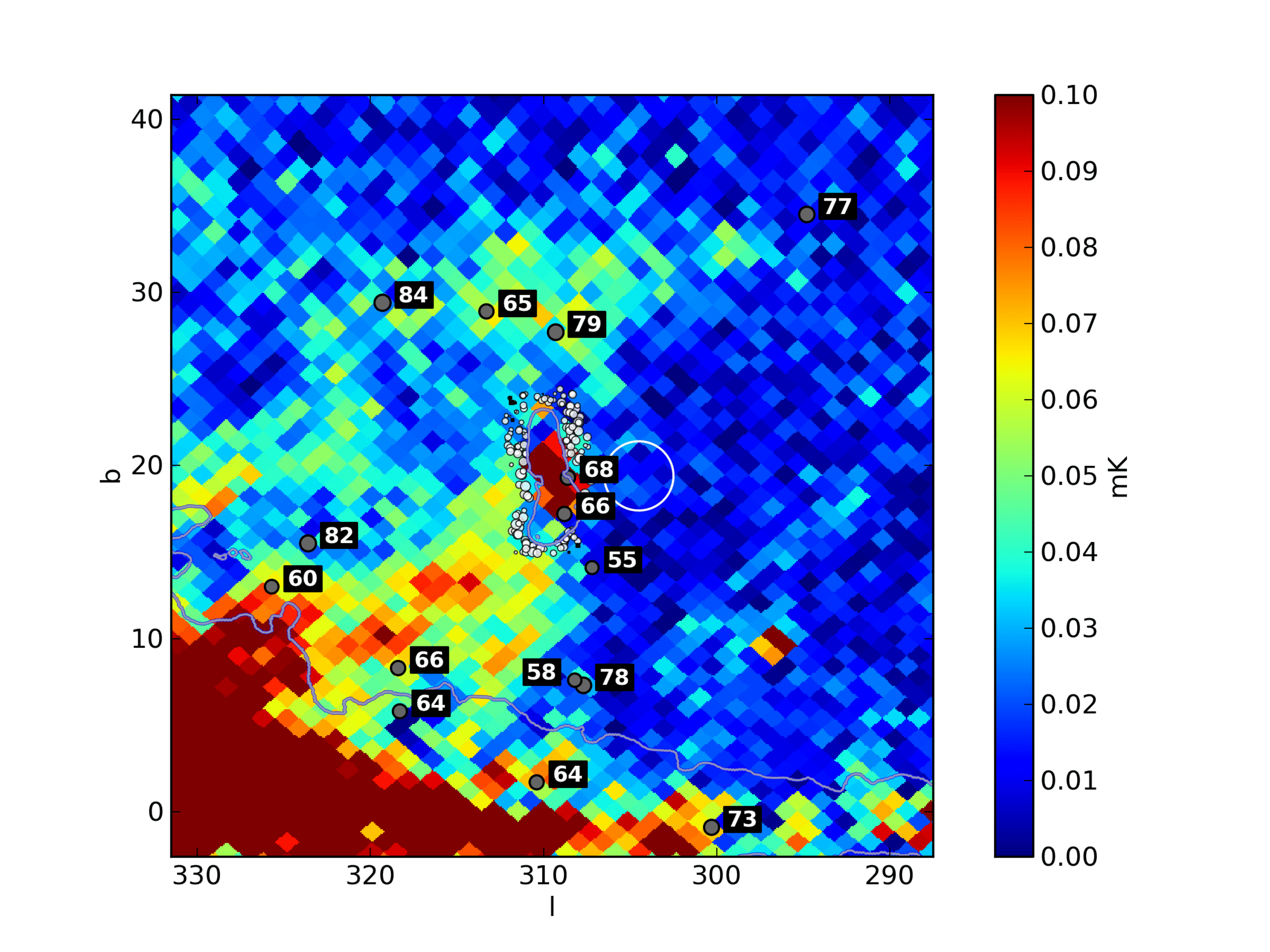}
\caption{Polarized synchrotron radiation at 22 GHz (color) from WMAP7 data \cite{Gold:2011}. The published Auger UHECR events above 55 EeV \cite{Auger:2010} around Cen A are indicated with small gray circles; their energies are given in EeV. Contours from radio data \cite{Haslam:1982} outline Cen A (center) and parts of the Galactic plane (both contours drawn at 70 K at 408 MHz). 160 extragalactic sources with line-of-sights outside Cen A \cite{Feain:2009} are shown along the boundary of Cen A. Filled white circles denote negative rotation measures (corresponding to a line-of-sight electron-density-weighted average magnetic field \emph{away} from the observer), black squares (very few) denote positive rotation measures. The size of the markers is proportional to the magnitude of the rotation measure. The large white circle shows the region 5$^\circ$ to the right, used to estimate the PI in the direction of Cen A without foreground contamination.}\label{cenA_data}
\end{figure}

Fig. \ref{cenA_data} shows a portion of the sky centered on Cen A, with color scale indicating the synchrotron polarization intensity (PI) from WMAP7. It can be seen  from Fig. \ref{cenA_data} that Cen A is located at the edge of a highly polarized region, part of the nearby North Polar Spur (NPS) or radio loop 1\cite{Wolleben:2007}.  If the measured PI near Cen A is dominated by these local structures, the emission is likely coming from very nearby ($\lesssim 200$ pc in the case of the NPS), in which case this region is not a good indicator of the large scale magnetic field relevant for UHECR deflection.  The measured PI in the direction of Cen A is likely dominated by emission from Cen A itself, and also cannot be used. But the area immediately to the right of Cen A  appears mostly uncontaminated by local emission.  We select the 2\degree\ radius disk shown in Fig. \ref{cenA_data}, centered on $l=304.5^\circ, \, b=+19.4^\circ$, a point 5 degrees from Cen A;  this point was chosen to be close to the direction of Cen A yet avoid any obvious contamination due to foreground sources.  Here we compute the average PI and I and their variance.  The PI is so small that the polarization angle cannot be reliably determined and we do not use it.  In this way we estimate the polarized intensity and intensity at 22 GHz in the general direction of Cen A to be $\mathrm{PI}=0.008\pm0.006$ mK and $\mathrm{I}=0.14\pm0.07$ mK, where the uncertainties quoted are the standard deviation of the 1$^\circ$x1$^\circ$ subpixels.   

The rotation measures of 160 extragalactic sources (EGS) with lines-of-sight near, but outside, Cen A were measured using the Australian Telescope Compact Array by Feain et al. \cite{Feain:2009} (see Fig. \ref{cenA_data}). Nearby foreground structures do not disproportionately impact RMs as they do for synchrotron emission, so these RMs provide an excellent measurement of the RM in this direction.  The average RM of these 160 new measurements is $-54\, \text{rad}\,\text{m}^{-2}$, with a standard deviation of $32\, \text{rad}\,\text{m}^{-2}$.  

The corresponding predictions of the combined JF12 coherent and random GMF models are $\mathrm{PI}=0.00895$ mK, $\mathrm{I}=0.17$ mK and $\mathrm{RM}=-51.092\,\text{rad}\,\text{m}^{-2}$.   Although the JF12 model parameters were determined using all available RMs including those of Feian et al.\cite{Feain:2009}, the fitting procedure reduces the 160 RMs surrounding Cen A to just a handful of data points (4\degree-by-4\degree\ pixels) out of several thousands used to constrain the GMF parameters, so there is no {\em a priori} guarantee that the model prediction will agree well with the local observables.   Thus the virtually perfect agreement between the JF12 model predictions and the observations along the Cen A line of sight, is a testimonial to the quality of the functional form of the model and the power of the global fitting approach.

The uncertainty on the mean is $1/\sqrt{N}$ times the standard deviation, where $N$ is the number of statistically independent measurements.  The extent to which the different subpixels for PI and the individual RMs can be considered independent measurements depends on the scale length of the random fluctuations, which has not yet been established.  Even with $N$'s of 16 and 160, leading to observational mean values of $\mathrm{PI}=0.008 \pm 0.0015 , \, \mathrm{I}=0.14\pm0.025$ mK and $\mathrm{RM}= -54 \pm 3$, the JF12 predictions are in excellent agreement within the errors.   Given the much worse global fit of other GMF models to the data, it is not surprising that they also give a poorer fit in the Cen A direction with  \{$\mathrm{PI} \,({\rm mK}), \, \mathrm{RM}= \,(\text{rad}\,\text{m}^{-2})$\} predictions:  SR10 \{0.00656, \,-25.486\}, P+11BSS \{0.0092, \,  -72.411\}, P+11ASS \{0.0096, \, -100.469\} , Stanev BSS \{0.00426, \, -56.594\}, Stanev ASS \{0.00426, \, 61.704\}.  

\begin{figure}
\centering
\includegraphics[width=1\linewidth]{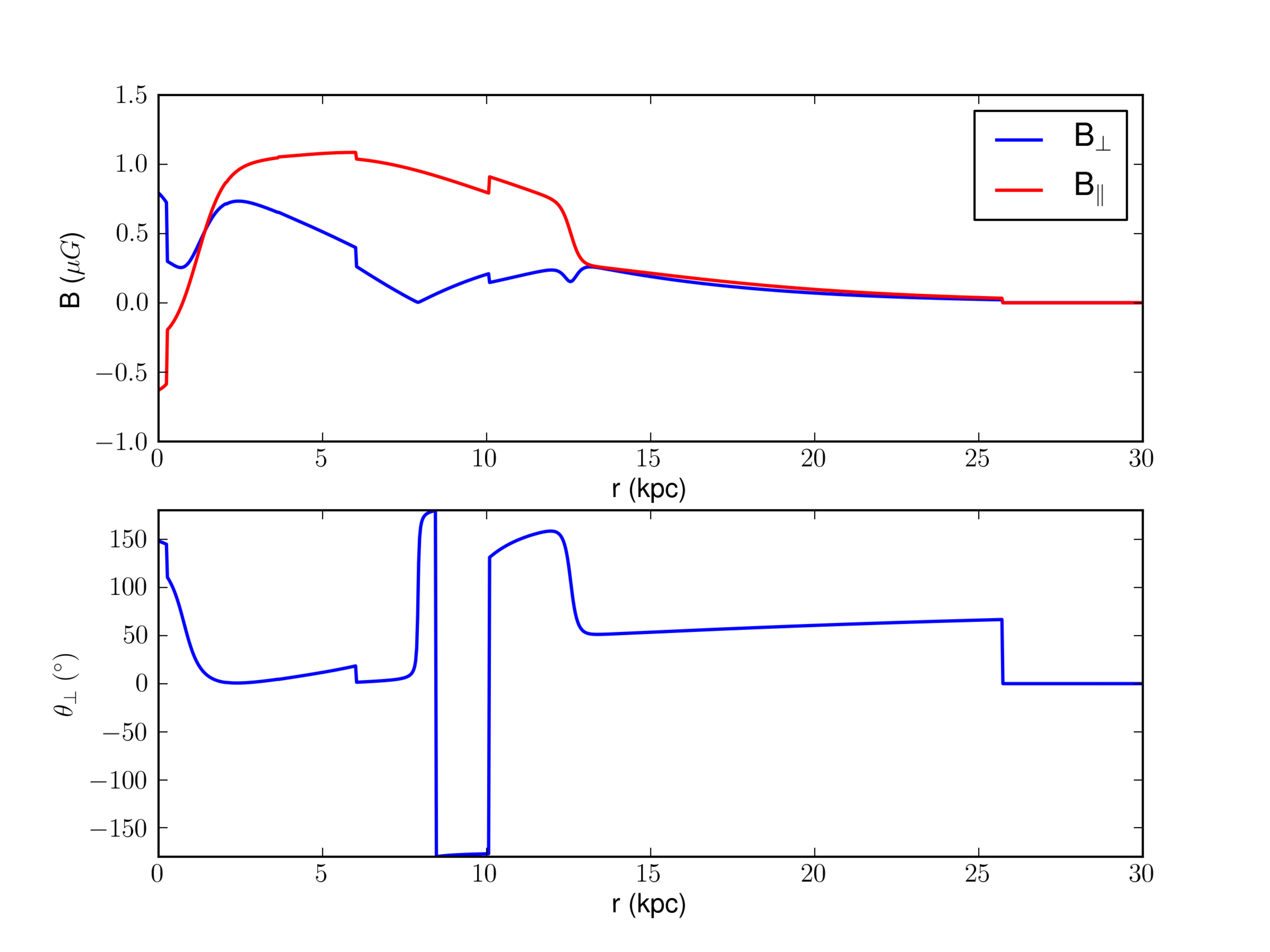}
\caption{The coherent field along the trajectory from Cen A, in the JF12 model.}\label{B_comp_CenA}
\end{figure}

One might think that polarized synchrotron data could be used directly to determine the deflections of UHECRs along a line-of-sight of interest, since both depend on the components of $\vec{B}$ transverse to the line of sight, but this is not the case for several reasons.  Most importantly, the Stokes parameters $Q$ and $U$ are sensitive only to the orientation and strength of the transverse field, but not its sign, whereas a UHECR's deflection depends on the sign of $\vec{B}_{\perp}$:  a trajectory which traverses a region with a sign reversal has a smaller net deflection than in the absence of the reversal.  In addition, $Q$ and $U$ are weighted by the relativistic electron density, $n_{{\rm cre}}$, making them much less sensitive to the field in the halo than are UHECR deflections.  (This means, of course, that when UHECR sources and charge assignments are known, UHECR deflections will become a very important tool for constraining the GMF.)   Therefore it is a necessary but not sufficient condition that a GMF model gives a good fit to the PI data along its trajectory:  a good {\em global} fit to the observables is also necessary to properly constrain the sign reversals.   Fig. \ref{B_comp_CenA} shows the coherent field along the line of sight to Cen A in the JF12 GMF model;  $\theta=0$ is angle of $\vec{B}_{\perp}$ with respect to the Galactic plane, in a right-handed coordinate system with  $\hat{z}$ pointing in the direction of motion.

\section{Deflections of UHECRs from Cen A}\label{deflections}

If there were no random component to the deflections of UHECRs, the arrival directions of UHECRs from any individual source would fall on a single arc of width the angular resolution of the Auger events ($\approx 1^\circ$), with the separation from the source region inversely proportional to the rigidity of the cosmic ray ($E/Z=$ \mbox{energy/charge}), in the limit of small-angle deflections.  This distribution of arrival directions is manifestly not observed in the Cen A region. However, if coherent fields are negligible compared to random ones, the arrival directions would be smeared over all directions rather than aligned.  Thus, the viability of the proposal that Cen A is the source of Auger's reported excess of UHECRs within $18^{\circ}$ with respect to the isotropic expectation,  depends on the relative importance of coherent and random deflections and the magnitude of the latter.  Therefore it is crucial to our analysis that the random fields along the trajectories of interest are well constrained as has been done by JF12 \cite{jf12rand}.  The effect of Galactic random fields on UHECR deflections was first considered in  \cite{Tinyakov:2005}; with the better-constrained JF12 random field, especially in the Cen A direction, the uncertainty in the random deflections is reduced.

JF12 also considered the possibility of striated fields -- a field with a definite orientation but having a random sign and magnitude, as could be produced by stretching a random field.   While random fields cause dispersion in all directions on the sky, striated fields aligned with the local regular field merely rescale the displacement due to the coherent field, similarly to UHECR energy resolution.  Using RM, $Q$ and $U$ alone, the existence of a striated random field aligned with the local coherent field cannot be distinguished from a rescaling of the relativistic electron density, $n_{\rm cre}$, and \cite{jf12} only constrained the product.  The degeneracy can be broken by fitting the total intensity as is done in ref. \cite{jf12rand}; we incorporate the effect of the striated random field along with those of the normal, isotropic random field, in the deflection calculations below. 

\subsection{Deflections in the Galactic magnetic field}

\begin{figure}
\centering
\includegraphics[width=1\linewidth]{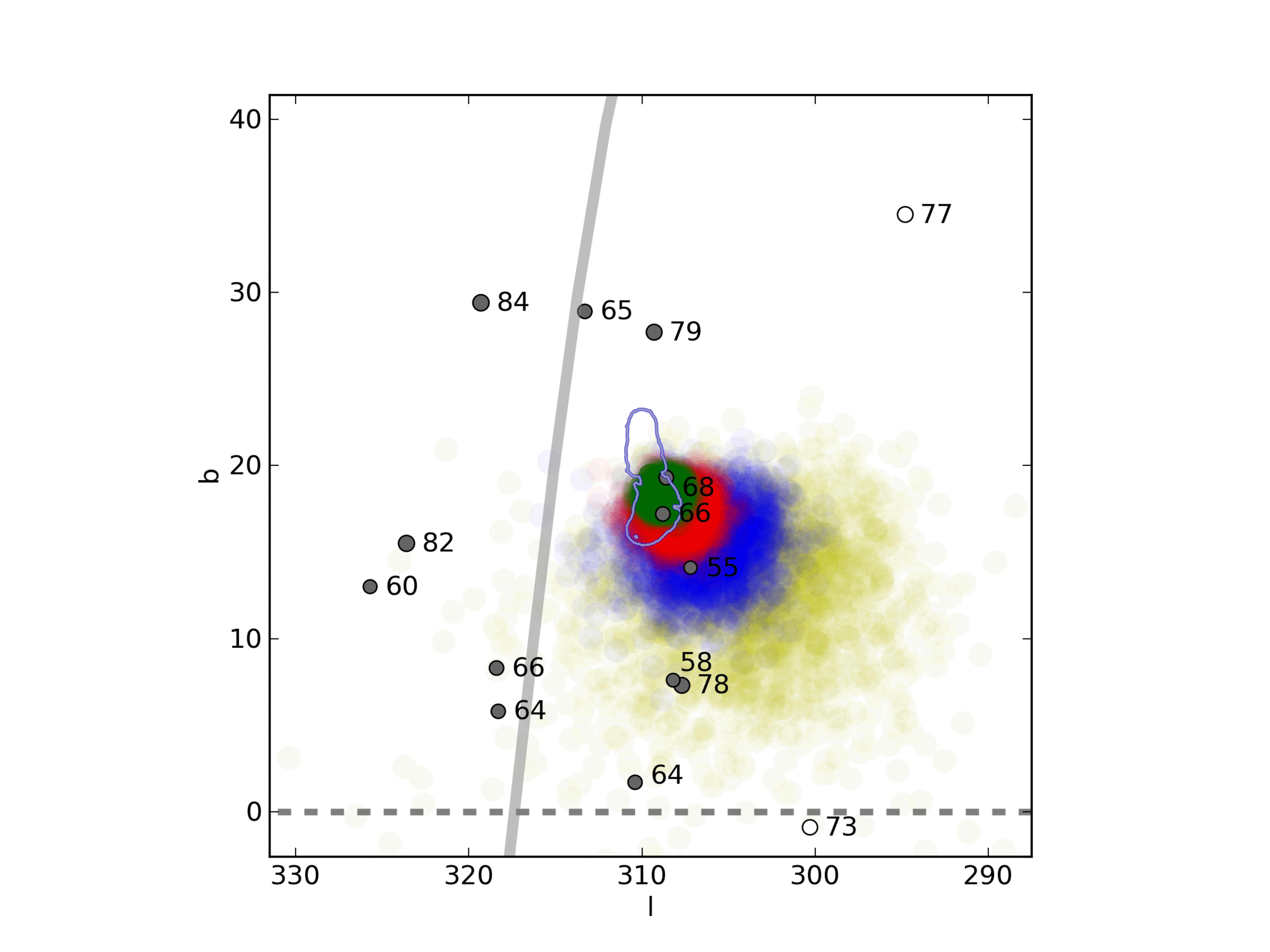}
\caption{Colored regions indicate the locus within which UHECRs should be observed if the source is the core of Cen A; green, red, blue and ochre regions are for UHECRs of rigidity 160, 80, 40 and 20 EeV/Z, respectively;  the super-Galactic plane is shown as a grey line; nearby UHECRs and their energies are indicated, those with open circles are more than 18$^{\circ}$ from Cen A.  The dispersion due to the random (turbulent) component of the magnetic field and observational uncertainties are both included in producing these probability distributions.    Note that the actual arrival directions will fall within a much narrower region within this domain because the trajectories of UHECRs of similar energies probe approximately the same random field.   The uncertainty in the mean deflection from the regular field, due to 1-sigma uncertainties in the GMF parameters, is shown in Fig. \ref{stanev}. }\label{rainbow}
\end{figure}

The deflection angle is inversely proportional to the cosmic ray rigidity, for small deflections. Using the JF12 best-fit global GMF model, we find that the magnitude of the arrival direction deflection, of a UHECR from the direction of Cen A in the regular magnetic field, is 
\beq
\label{regdef}
\Delta\theta_{\rm reg} = ( 2.3^\circ\pm 0.24 ^\circ) ~ (Z/E_{100}),
\end{equation}
valid for energies for which this is a small deflection.  Here, $Z$ is the charge in units of the proton charge, $E_{100}$ is the energy of the UHECR in units of 100 EeV, and $0.24^\circ$ is the uncertainty due to GMF parameter uncertainties (the standard deviation of the separation in arrival directions relative to that of the best-fit-field, using GMFs for randomly chosen parameter sets drawn from the Markov Chain Monte Carlo probability distribution of JF12 parameters).  The CR is deflected so that it arrives from a direction closer to the Galactic plane and somewhat farther from the Galactic center, as shown in Fig. \ref{rainbow}.

To estimate the locus of uncertainty in arrival direction due to random magnetic deflections we repeatedly propagate UHECRs of different rigidity through the Galaxy, dividing the path into domains of size $\lambda$.  The field in the $i$th domain is the sum of the JF12 coherent GMF, $\vec{B_i}$, plus a random part, $B_{{\rm rand},i} \hat{n}$, and a striated-random part $B_{{\rm stri},i} \hat{\eta}$, where $\hat{n}$ is a randomly oriented unit vector chosen to have a different random direction in each step and $\hat{\eta}$ is a unit vector oriented along the direction of the local coherent field with sign randomly chosen in each step.   $B_{{\rm rand},i}$ is the rms random field evaluated in the center of the $i$th domain using the JF12 random field model\cite{jf12rand}, and  $B_{{\rm stri},i}$ is $\sqrt{\beta} | B_{{\rm reg},i} | $, with $\beta = 1.38$\cite{jf12,jf12rand}.  The domain size $\lambda$ corresponds to the maximum coherence length of the turbulent field; this is uncertain but is plausibly of order $\approx$ 100 pc, a typical maximum size of supernova remnants\cite{Gaensler:1995, Haverkorn:2008}.  As one would expect, the spread in arrival directions $\propto \sqrt{\lambda}$.  For UHECRs arriving from Cen A the mean angular separation in 1000 different realizations of the random field, between the arrival direction and the centroid of the arrival-direction distribution, is 
\beq \label{sigrand}
\sigma_{\rm rand} = 1.3^\circ  (Z/E_{100}) \sqrt{ \lambda_{100} },
\end{equation} 
where $\lambda_{100}=\lambda/100$ pc.  A more sophisticated treatment of the random field realization with a Kolmogorov spectrum would reduce the dispersion compared to Eq. \ref{sigrand}, since in that case power is shared over a range of scales rather than being concentrated in the largest coherence length which is most effective at deflecting UHECRs.   

It is important to emphasize that the dispersion quoted in Eq. \ref{sigrand} and  the colored regions shown in Fig. \ref{rainbow} {\em do not} represent the typical spread of arrival directions predicted for UHECRs coming from Cen A, because UHECRs of a given energy from a given source direction and entry point into the GMF follow nearly the same trajectories and thus propagate in nearly the same fields for much of their trajectories, whereas the calculation above takes different fields for each UHECR to obtain the inclusive region of possible arrival directions, given our knowledge of the GMF.  
We can gain insight into whether different events of the same energy probe different coherence regions as follows.  The deflection of an individual UHECR in a given random magnetic field can be described as a random walk with net deflection $\sqrt{N}$ times the deflection $\delta$ due to the random field in a typical coherence length: $\sigma_{{\rm rand}} \approx \sqrt{N} \, \delta$, where $\sqrt{N}$ is the number of independent random domains.  From Fig. \ref{B_comp_CenA} and taking a 100 pc coherence length, $\sqrt{N} \approx 12$ crossing the Galaxy en route from Cen A.  With $\sigma_{{\rm rand}} \approx 1.3^{\circ}$ we infer $\delta \approx 0.1^{\circ}$, so the lateral displacement relative to the trajectory in the coherent field alone, in crossing a typical coherence region, is $\approx 0.2$ pc.  This is far smaller than the assumed 100 pc size of the coherence region, so random UHECR deflections do not cause trajectories for events of the same energy to diverge enough to sample different random magnetic field domains.  More study is needed to determine whether the divergence of the beam from Cen A or the lateral size of the portion of the beam focussed on Earth, may be sufficient for different events to probe different magnetic domains.  Thus Eq. (\ref{sigrand}) is an upper bound on the magnetic dispersion from random Galactic fields in the JF12 model.

\subsection{Deflections in extragalactic magnetic fields}

The direction of arrival to the Galaxy is also smeared compared to the direction of the source, by deflections in the Cen A system itself and by the extragalactic magnetic field (EGMF) between Cen A and the Milky Way.  But even very large deflections within the Cen A region can only move the apparent source direction within the radio lobes, translating the image region accordingly.  Thus we are concerned with the EGMF.

The EGMF is thought to be turbulent, with a field strength typically assumed to be of order 0.1 - 1 nG \cite{Kronberg:1994} and coherence length $\lambda_{\rm EG} \approx 100$ kpc.  It was pointed out in \cite{Farrar:2000b} that the EGMF of the low-redshift universe is poorly constrained by the dispersions in RM of high redshift sources as was done in the classic work of \cite{Kronberg:1994}, and also that \cite{Kronberg:1994} uses an unrealistically large value for the electron density and thus underestimates the true field.   Refs. \cite{Neronov:2009,neronovVovkSci10,dermer+blazarEGMF11,TaylorVockNeronovEGMF11} provide an overview of recent efforts to constrain the EGMF using TeV blazers.

If UHECRs experience many small deflections over a distance $D$ from source to Galaxy, in a turbulent EGMF whose rms value and coherence length are $B_{\rm EG}$ and $\lambda_{{\rm EG}}$, then the rms angle between the source and the UHECR arrival direction to the Galaxy is \cite{Waxman:1996}
\beq
\label{egmfdef}
\delta \theta_{\rm EG} \approx 0.15^\circ \, \, \left( \frac{D}{3.8 \, \rm Mpc} \cdot \frac{\lambda_{\rm EG} }{100 \,\rm kpc } \right)^{\frac{1}{2} }\, (B_{\rm EG}/1\, {\rm nG}) \,\, (Z/E_{100}),
\end{equation}
assuming the lateral extent of the EGMF is $\gtrsim  D \, \delta \theta $.  The extragalactic random deflection predicted in Eq. (\ref{egmfdef}) must be added in quadrature with the Galactic random deflection given in Eq. (\ref{sigrand}) to obtain the total random deflection.  
In order for $\delta \theta_{\rm EG} \geq 12^\circ$ (or just to be greater than the coherent Galactic deflection in Eq. \ref{regdef}), we would need
\beq
\label{egmfMin}
B_{\rm EG} \gtrsim  80 \, (10) \, \left(\frac{\lambda_{\rm EG} }{100 \,\rm kpc } \right)^{-\frac{1}{2} } {\rm nG}.
\end{equation}

Deflections in a variety of realizations of  extragalactic fields between Cen A and the edge of the Galaxy were recently considered in \cite{yuksel+CenA12}.  They attempted to find choices of rms strength and maximum coherence length of the EGMF in the intervening region, which could produce the observed distribution of UHECR arrival directions within $18^\circ$ of Cen A, assuming that Cen A is the source and ignoring GMF deflections.   They were interested in the possibility that smaller rms field strength and larger coherence length than we employed above, could generate the observed pattern.  However they find that except for specially selected examples, EGMFs do not reproduce the combination of small mean deflection and large rms deflection seen in the Auger events within $18^\circ$ of Cen A, taking Cen A to be the source.  

\subsection{Futuristic Note: RMs of Cen A sources and UHECR deflections} 

In the small-angle deflection approximation one can directly relate the smearing in UHECR arrival directions to the dispersion in RMs accumulated in any given region in which the electron density is roughly constant. If the magnetic field is turbulent with a typical coherence length $\lambda$ and rms strength $B$, over a region of size $R = N \, \lambda$, the rms dispersion in RM is (in $ {\rm rad \, m^{-2}}$)
\beq
\label{deltaRM}
\delta_{\rm RM } = \sqrt{N} \, 8.1\times 10^4 \,\, n_e \, \frac{ B_{\mu \rm G}}{ \sqrt{3}} \, \frac{\lambda}{100 \,\rm kpc},
\end{equation}
where $n_e$ is the electron density in ${\rm cm}^{-3}$.  A UHECR of energy $E_{100}$ EeV and charge $Z$ propagating from the center of such a region would typically have an angular deflection
\beq
\label{deltatheta}
\delta \, \theta =  \sqrt{\frac{N}{2}} \sqrt{\frac{2}{3}}\, \frac{ \lambda }{ R_{\rm Larmor}} = 0.52 \times\sqrt{N} \, \lambda_{100} \left(\frac{B_{\mu \rm G} \,Z}{E_{100}}\right),
\end{equation}
where the factor $\sqrt{2/3}$ is due to the projection of the Larmor radius to the plane transverse to the line-of-sight. Combining Eq. (\ref{deltaRM}) and (\ref{deltatheta}) we obtain
\beq
\label{dispRMdispCR}
\delta \, \theta_{\rm EG} \approx (6.4\times10^{-4})^\circ \, \delta_{\rm RM }\,(Z/E_{100})\, n_{e}^{-1}.
\end{equation}
Unfortunately, in spite of the greatly improved data on RMs near Cen A \cite{Feain:2009}, Eq. (\ref{dispRMdispCR}) does not provide a useful constraint on the extragalactic random field between Cen A and the Milky Way, because the Galaxy's much larger electron density and field strengths means that Faraday rotation in Galactic fields dominates the RM accumulated between Cen A and the Galaxy.  When the GMF and electron densities are much better known, Eq. (\ref{dispRMdispCR}) could in principle become useful to eliminate the sensitivity of the random deflection prediction to the uncertain coherence length and field strength. 


\section{Cen A as the source of proton UHECRs}

In the JF12 GMF model, a 60 EeV proton is deflected $\approx 3.8^\circ$ toward the Galactic plane, at an angle of about 45$^\circ$ away from the Galactic center.  This prediction has an rms uncertainty of $\approx  \frac{1}{4}^\circ$ about the mean deflection for the best-fit GMF parameters, coming from the uncertainty in GMF parameters; the locus is shown in Fig. \ref{stanev}.  Smearing due to random magnetic deflections is sub-dominant by at least a factor of two compared to the deflection due to the regular field, as can be seen in Fig. \ref{rainbow} which shows the maximal arrival locus of UHECRs of rigidities 160, 80, 40 and 20 originating in the Cen A galaxy.  The Auger collaboration estimates a $\approx 14\%$ statistical and $\approx 22\%$ systematic uncertainty on the UHECR energies \cite{Auger_energy:2009}.  Adding them in quadrature produces a $\approx 26\%$ energy uncertainty for each event.  Since deflections $\sim 1/E$, this produces an additional uncertainty factor in the overall deflection ranging from $0.8-1.35$.

Taking a generous view of the random deflections and uncertainties in the GMF, it appears that up to three of the 69 published Auger UHECRs above 55 EeV could be protons originating from Cen A or its radio lobes.  Two to three more events could be consistent with a Cen A origin, taking into account that the GMF model is probably not perfect, if their charges are $Z \sim 2-4$; events with higher charges would be deflected more than $18^{\circ}$.

The Galactic magnetic field deflects UHECRs coming from Cen A or its radio lobes into a swath, with events of lower rigidity further from the source. Fig. \ref{rainbow} shows the expected swath of UHECRs for rigidities, E/Z = 20, 40, 80, and 160 EV if Cen A itself is the source. The opening angle of the region containing the UHECRs is given by  $\sigma_{\rm rand}/|\Delta\theta_{\rm reg}| \lesssim 35\degree$.  As long as E/Z is small enough that $\sigma_{\rm rand}$ and $\Delta \theta_{\rm reg}$ do not become large, the numerator and denominator in this expression are both inversely proportional to the rigidity of the cosmic rays. Thus the opening angle of the swath is \emph{independent} of the energy calibration, charge, or composition, of the UHECRs.  

The maximum opening angle of the arrival-direction-swath for UHECRs originating in Cen A is not only independent of uncertainties about the UHECR, it is a particularly robust and reliable prediction of the JF12 model.  This is because the largest source of uncertainty in the JF12 work is its dependence on $n_{{\rm cre}}(\vec{r})$ which is based on GALPROP modeling\cite{jf12,jf12rand}.  But $n_{{\rm cre}}(\vec{r})$ is a factor common to all of the observables $I$, $Q$, and $U$ constraining respectively the random field and transverse coherent field, so the relative strength of the coherent and random deflections and hence opening angle of the swath, is quite robust.  Details of the coherent field will change as $n_{{\rm cre}}$ and the functional form of the field model are refined, resulting in possible rotation of the orientation of the arrival-direction-swath.  However the general relationship between magnitudes of random and coherent deflections is likely to persist  as noted, so the swath would have a roughly similar appearance, but possibly be pivoted about Cen A by some amount and/or the footprint be slightly stretched or squished.  

Hence, it is unlikely that more than 5 to 6 of the 13 Auger UHECRs within 18$^\circ$ in the published 69 event data-set can be attributed to Cen A or its radio lobes, in spite of uncertainties in the composition or energy calibration.  We emphasize that the arrival direction locus shown in Fig. \ref{rainbow} indicates the probability distribution for arrival directions for the given random field:  it does not represent the actual image of Cen A in UHECRs, because the actual GMF is a particular realization of the random ensemble used to produce Fig. \ref{rainbow}.  The actual arrival directions should fall in a considerably narrower band, whose width depends on the spatial properties of the random field, especially its maximum coherence length, which are not yet determined.  It is tantalizing that 5-6 events do fall on such a narrow arc within the predicted Cen A locus.

\section{Summary}

We have checked that the new JF12 model of the Galactic magnetic field \cite{jf12}, which gives a very good global fit to a large amount of Faraday Rotation Measure and synchrotron emission data, also gives an accurate accounting of the extensive RM data as well as the synchrotron data in the particular direction of Cen A.  This justifies confidence in the predictions of the JF12 GMF model for UHECR deflections.  

Using the JF12 model of the large scale GMF\cite{jf12} and the random and striated fields\cite{jf12rand}, we determine the locus in which protons and low-Z nuclei from Cen A should be found (Fig. \ref{rainbow}).  Three UHECRs in the published Auger events above 55 EeV (those closest to Cen A), have arrival directions consistent with their being protons which originated in Cen A or its radio lobes.   Three more events with energies 58, 78 and 64 EeV fall in the arrival locus for rigidity $E/Z \leq 20\,$EV originating from Cen A; they could have $Z=2-4$ and originate in Cen A.  Events with higher charges and $E \lesssim 60 $ EeV are deflected more than $18^{\circ}$.  Extragalactic fields would have to be $\sim 80$ nG -- far stronger than conventionally assumed -- in order for most UHECRs within $18^\circ$ of Cen A to have been produced by Cen A.

\acknowledgments

The research of R.J.\ and G.R.F.\  has been supported in part by NSF-PHY-0701451 and NASA grant NNX10AC96G.  
B.M.G.\ acknowledges the support of a Federation Fellowship from the Australian Research Council through grant FF0561298. The Australia Telescope Compact Array is part of the Australia Telescope which is funded by the Commonwealth of Australia for operation as a National Facility managed by CSIRO.  
G.R.F.\ acknowledges her membership in the Pierre Auger Collaboration and thanks her colleagues for their support of and contributions to her research. 



\begin{thebibliography}{99}
\bibitem{Greisen:1966}
K.~{Greisen}, {\it {End to the Cosmic-Ray Spectrum?}},  {\em Physical Review
  Letters} {\bf 16} (Apr., 1966) 748--750.

\bibitem{Zatsepin:1966}
G.~T. {Zatsepin} and V.~A. {Kuz'min}, {\it {Upper Limit of the Spectrum of
  Cosmic Rays}},  {\em Soviet Journal of Experimental and Theoretical Physics
  Letters} {\bf 4} (Aug., 1966) 78.

\bibitem{Cavallo:1978}
G.~{Cavallo}, {\it {On the sources of ultra-high energy cosmic rays}},  {\em
  A{\&}A} {\bf 65} (May, 1978) 415--419.

\bibitem{Romero:1996}
G.~E. {Romero}, J.~A. {Combi}, S.~E. {Perez Bergliaffa}, and L.~A.
  {Anchordoqui}, {\it {Centaurus A as a source of extragalactic cosmic rays
  with arrival energies well beyond the GZK cutoff}},  {\em Astroparticle
  Physics} {\bf 5} (Oct., 1996) 279--283,
  [\href{http://arxiv.org/abs/arXiv:gr-qc/9511031}{{\tt arXiv:gr-qc/9511031}}].

\bibitem{Farrar:2000}
G.~R. {Farrar} and T.~{Piran}, {\it {Deducing the Source of Ultrahigh Energy
  Cosmic Rays}},  {\em ArXiv: astro-ph/0010370} (Oct., 2000)
  [\href{http://arxiv.org/abs/arXiv: astro-ph/0010370}{{\tt arXiv:
  astro-ph/0010370}}].

\bibitem{Auger:2010}
{The Pierre AUGER Collaboration}, P.~{Abreu}, M.~{Aglietta}, E.~J. {Ahn},
  D.~{Allard}, I.~{Allekotte}, J.~{Allen}, J.~{Alvarez Castillo},
  J.~{Alvarez-Mu{\~n}iz}, M.~{Ambrosio}, and et~al., {\it {Update on the
  correlation of the highest energy cosmic rays with nearby extragalactic
  matter}},  {\em Astroparticle Physics} {\bf 34} (Dec., 2010) 314--326,
  [\href{http://arxiv.org/abs/1009.1855}{{\tt arXiv:1009.1855}}].

\bibitem{Wibig:2007}
T.~{Wibig} and A.~W. {Wolfendale}, {\it {Heavy Cosmic Ray Nuclei from
  Extragalactic Sources above 'The Ankle'}},  {\em {OAJ}} {\bf 2} (Dec., 2007)
  95--101, [\href{http://arxiv.org/abs/0712.3403}{{\tt arXiv:0712.3403}}].

\bibitem{Gorbunov:2008}
D.~S. {Gorbunov}, P.~G. {Tinyakov}, I.~I. {Tkachev}, and S.~V. {Troitsky}, {\it
  {On the interpretation of the cosmic-ray anisotropy at ultra-high energies}},
   \href{http://arxiv.org/abs/0804.1088}{{\tt arXiv:0804.1088}}.

\bibitem{Hardcastle:2009}
M.~J. {Hardcastle}, C.~C. {Cheung}, I.~J. {Feain}, and {\L}.~{Stawarz}, {\it
  {High-energy particle acceleration and production of ultra-high-energy cosmic
  rays in the giant lobes of Centaurus A}},  {\em MNRAS} {\bf 393} (Mar.,
  2009) 1041--1053, [\href{http://arxiv.org/abs/0808.1593}{{\tt
  arXiv:0808.1593}}].

\bibitem{Kachelriess:2009}
M.~{Kachelrie{\ss}}, S.~{Ostapchenko}, and R.~{Tom{\`a}s}, {\it {High energy
  radiation from Centaurus A}},  {\em New Journal of Physics} {\bf 11} (June,
  2009) 065017, [\href{http://arxiv.org/abs/0805.2608}{{\tt arXiv:0805.2608}}].

\bibitem{Rachen:2008}
J.~P. {Rachen}, {\it {Ultra-high energy cosmic rays from radio galaxies
  revisited}},  in {\em XXth Rencontres de Blois}, Challenges in Particle
  Astrophysics, Aug., 2008.

\bibitem{Fargion:2008}
D.~{Fargion}, {\it {Light nuclei solving the AUGER puzzles: the Cen-A
  imprint}},  {\em Physica Scripta} {\bf 78} (Oct., 2008) 045901,
  [\href{http://arxiv.org/abs/0801.0227}{{\tt arXiv:0801.0227}}].

\bibitem{Fargion:2009}
D.~{Fargion}, {\it {Coherent and random UHECR Spectroscopy of Lightest Nuclei
  along CenA: Shadows on GZK Tau Neutrinos spread in a near sky and time}},
  {\em ArXiv e-prints} (Aug., 2009) [\href{http://arxiv.org/abs/0908.2650}{{\tt
  arXiv:0908.2650}}].

\bibitem{AugerICRCcorrelations:2009}
J.~Abraham {\em et~al.}, {\it {Astrophysical Sources of Cosmic Rays and Related
  Measurements with the Pierre Auger Observatory}},  {\em 31st ICRC, Lodz,
  Poland} (June, 2009) [\href{http://arxiv.org/abs/0906.2347}{{\tt
  arXiv:0906.2347}}].

\bibitem{AugerAGN:2007}
{\bf Pierre Auger} Collaboration, J.~Abraham {\em et~al.}, {\it {Correlation of
  the highest energy cosmic rays with nearby extragalactic objects}},  {\em
  Science} {\bf 318} (2007) 938--943,
  [\href{http://arxiv.org/abs/0711.2256}{{\tt arXiv:0711.2256}}].

\bibitem{AugerICRCcomposition:2009}
J.~Abraham {\em et~al.}, {\it {Studies of Cosmic Ray Composition and Air Shower
  Structure with the Pierre Auger Observatory}},  {\em Proceedings of ICRC
  2009, Lodz, Poland} (June, 2009) [\href{http://arxiv.org/abs/0906.2319}{{\tt
  arXiv:0906.2319}}].

\bibitem{jf12}
R.~{Jansson} and G.~R. {Farrar}, {\it {A New Model of the Galactic Magnetic
  Field}}, Physica Scripta {\bf 757} (2012)
  [\href{http://arxiv.org/abs/1204.3662}{{\tt arXiv:1204.3662}}].

\bibitem{jf12rand}
R.~{Jansson} and G.~R. {Farrar}, {\it {The Galactic Magnetic Field}},  {\em Ap.J. Lett. in press}.

\bibitem{Sun:2008}
X.~H. {Sun}, W.~{Reich}, A.~{Waelkens}, and T.~A. {En{\ss}lin}, {\it {Radio
  observational constraints on Galactic 3D-emission models}},  {\em A{\&}A} {\bf
  477} (Jan., 2008) 573--592, [\href{http://arxiv.org/abs/arXiv:0711.1572}{{\tt
  arXiv:0711.1572}}].

\bibitem{Sun:2010}
X.-H. {Sun} and W.~{Reich}, {\it {The Galactic halo magnetic field revisited}},
   {\em Research in Astronomy and Astrophysics} {\bf 10} (Dec., 2010)
  1287--1297, [\href{http://arxiv.org/abs/1010.4394}{{\tt arXiv:1010.4394}}].

\bibitem{pshirkov+11}
M.~S. {Pshirkov}, P.~G. {Tinyakov}, P.~P. {Kronberg}, and K.~J. {Newton-McGee},
  {\it {Deriving the Global Structure of the Galactic Magnetic Field from
  Faraday Rotation Measures of Extragalactic Sources}},  {\em Ap. J. } {\bf 738}
  (Sept., 2011) 192, [\href{http://arxiv.org/abs/1103.0814}{{\tt
  arXiv:1103.0814}}].

\bibitem{Prouza2003}
M.~Prouza and R.~Smida, {\it The galactic magnetic field and propagation of
  ultra-high energy cosmic rays},  {\em Astron. Astrophys.} {\bf 410} (2003)
  1--10.

\bibitem{Stanev:1997}
T.~{Stanev}, {\it {Ultra--High-Energy Cosmic Rays and the Large-Scale Structure
  of the Galactic Magnetic Field}},  {\em Ap. J. } {\bf 479} (Apr., 1997) 290,
  [\href{http://arxiv.org/abs/arXiv:astro-ph/9607086}{{\tt
  arXiv:astro-ph/9607086}}].

\bibitem{Takami:2009}
H.~{Takami} and K.~{Sato}, {\it {Does Galactic Magnetic Field Disturb the
  Correlation of the Highest Energy Cosmic Rays with Their Sources?}},  {\em
  Ap. J. } {\bf 724} (Dec., 2010) 1456--1472,
  [\href{http://arxiv.org/abs/0909.1532}{{\tt arXiv:0909.1532}}].

\bibitem{Vorobiov:2009}
S.~{Vorobiov}, M.~{Hussain}, and D.~{Veberi{\v c}}, {\it {Studies of UHECR
  propagation in the galactic magnetic field}},  {\em Nuclear Physics B
  Proceedings Supplements} {\bf 196} (Dec., 2009) 203--206,
  [\href{http://arxiv.org/abs/0902.3123}{{\tt arXiv:0902.3123}}].

\bibitem{Feain:2009}
I.~J. {Feain}, R.~D. {Ekers}, T.~{Murphy}, B.~M. {Gaensler}, J.~{Macquart},
  R.~P. {Norris}, T.~J. {Cornwell}, M.~{Johnston-Hollitt}, J.~{Ott}, and
  E.~{Middelberg}, {\it {Faraday Rotation Structure on Kiloparsec Scales in the
  Radio Lobes of Centaurus A}},  {\em Ap. J. } {\bf 707} (Dec., 2009) 114--125,
  [\href{http://arxiv.org/abs/0910.3458}{{\tt arXiv:0910.3458}}].

\bibitem{Israel:1998}
F.~P. {Israel}, {\it {Centaurus A - NGC 5128}},  {\em Astron. Astrophys. Rev.} {\bf 8} (1998)
  237--278, [\href{http://arxiv.org/abs/arXiv:astro-ph/9811051}{{\tt
  arXiv:astro-ph/9811051}}].

  \bibitem{Harris:2009}
G.~L.~H. {Harris}, M.~{Rejkuba}, and W.~E. {Harris}, {\it {The Distance to NGC
  5128 (Centaurus A)}},  {\em Pub. Ast. Soc. Aust.} {\bf 27} (Oct., 2010) 457--462,
  [\href{http://arxiv.org/abs/0911.3180}{{\tt arXiv:0911.3180}}].

\bibitem{Rieger:2009}
F.~M. {Rieger} and F.~A. {Aharonian}, {\it {Centaurus A as TeV {$\gamma$}-ray
  and possible UHE cosmic-ray source}},  {\em A{\&}A} {\bf 506} (Nov., 2009)
  L41--L44, [\href{http://arxiv.org/abs/0910.2327}{{\tt arXiv:0910.2327}}].

\bibitem{Moskalenko:2009}
I.~V. {Moskalenko}, L.~{Stawarz}, T.~A. {Porter}, and C.~C. {Cheung}, {\it {On
  the Possible Association of Ultra High Energy Cosmic Rays With Nearby Active
  Galaxies}},  {\em Ap. J. } {\bf 693} (Mar., 2009) 1261--1274,
  [\href{http://arxiv.org/abs/0805.1260}{{\tt arXiv:0805.1260}}].

\bibitem{Gold:2011}
B.~{Gold}, N.~{Odegard}, J.~L. {Weiland}, R.~S. {Hill}, A.~{Kogut}, C.~L.
  {Bennett}, G.~{Hinshaw}, X.~{Chen}, J.~{Dunkley}, M.~{Halpern}, N.~{Jarosik},
  E.~{Komatsu}, D.~{Larson}, M.~{Limon}, S.~S. {Meyer}, M.~R. {Nolta},
  L.~{Page}, K.~M. {Smith}, D.~N. {Spergel}, G.~S. {Tucker}, E.~{Wollack}, and
  E.~L. {Wright}, {\it {Seven-year Wilkinson Microwave Anisotropy Probe (WMAP)
  Observations: Galactic Foreground Emission}},  {\em Ap. J. Supp.} {\bf 192} (Feb.,
  2011) 15, [\href{http://arxiv.org/abs/1001.4555}{{\tt arXiv:1001.4555}}].

\bibitem{Wolleben:2007}
M.~{Wolleben}, {\it {A New Model for the Loop I (North Polar Spur) Region}},
  {\em Ap. J. } {\bf 664} (July, 2007) 349--356,
  [\href{http://arxiv.org/abs/0704.0276}{{\tt arXiv:0704.0276}}].

\bibitem{Haslam:1982}
C.~G.~T. {Haslam}, C.~J. {Salter}, H.~{Stoffel}, and W.~E. {Wilson}, {\it {A
  408 MHz all-sky continuum survey. II - The atlas of contour maps}},  {\em
  A{\&}A Supp.} {\bf 47} (Jan., 1982) 1.

\bibitem{Tinyakov:2005}
P.~G. {Tinyakov} and I.~I. {Tkachev}, {\it {Deflections of cosmic rays in a
  random component of the Galactic magnetic field}},  {\em Astroparticle
  Physics} {\bf 24} (Sept., 2005) 32--43,
  [\href{http://arxiv.org/abs/arXiv:astro-ph/0411669}{{\tt
  arXiv:astro-ph/0411669}}].

\bibitem{Gaensler:1995}
B.~M. {Gaensler} and S.~{Johnston}, {\it {The pulsar/supernova remnant
  connection}},  {\em MNRAS} {\bf 277} (Dec., 1995) 1243--1253.

\bibitem{Haverkorn:2008}
M.~{Haverkorn}, J.~C. {Brown}, B.~M. {Gaensler}, and N.~M. {McClure-Griffiths},
  {\it {The Outer Scale of Turbulence in the Magnetoionized Galactic
  Interstellar Medium}},  {\em Ap. J. } {\bf 680} (June, 2008) 362--370,
  [\href{http://arxiv.org/abs/0802.2740}{{\tt arXiv:0802.2740}}].

\bibitem{Kronberg:1994}
P.~P. {Kronberg}, {\it {Extragalactic magnetic fields}},  {\em Reports on
  Progress in Physics} {\bf 57} (Apr., 1994) 325--382.

\bibitem{Farrar:2000b}
G.~R. {Farrar} and T.~{Piran}, {\it {Violation of the Greisen-Zatsepin-Kuzmin
  Cutoff: A Tempest in a (Magnetic) Teapot? Why Cosmic Ray Energies above
  $10^{20}$ eV May Not Require New Physics}},  {\em Physical Review Letters}
  {\bf 84} (Apr., 2000) 3527--3530,
  [\href{http://arxiv.org/abs/arXiv:astro-ph/9906431}{{\tt
  arXiv:astro-ph/9906431}}].

\bibitem{Neronov:2009}
A.~{Neronov} and D.~V. {Semikoz}, {\it {Sensitivity of {$\gamma$}-ray
  telescopes for detection of magnetic fields in the intergalactic medium}},
  {\em Phys. Rev. D} {\bf 80} (Dec., 2009) 123012,
  [\href{http://arxiv.org/abs/0910.1920}{{\tt arXiv:0910.1920}}].

\bibitem{neronovVovkSci10}
A.~{Neronov} and I.~{Vovk}, {\it {Evidence for Strong Extragalactic Magnetic
  Fields from Fermi Observations of TeV Blazars}},  {\em Science} {\bf 328}
  (Apr., 2010) 73--, [\href{http://arxiv.org/abs/1006.3504}{{\tt
  arXiv:1006.3504}}].

\bibitem{dermer+blazarEGMF11}
C.~D. {Dermer}, M.~{Cavadini}, S.~{Razzaque}, J.~D. {Finke}, J.~{Chiang}, and
  B.~{Lott}, {\it {Time Delay of Cascade Radiation for TeV Blazars and the
  Measurement of the Intergalactic Magnetic Field}},  {\em Ap. J. Lett.} {\bf 733}
  (June, 2011) L21, [\href{http://arxiv.org/abs/1011.6660}{{\tt
  arXiv:1011.6660}}].

\bibitem{TaylorVockNeronovEGMF11}
A.~M. {Taylor}, I.~{Vovk}, and A.~{Neronov}, {\it {Extragalactic magnetic
  fields constraints from simultaneous GeV-TeV observations of blazars}},  {\em
  A{\&}A} {\bf 529} (May, 2011) A144, [\href{http://arxiv.org/abs/1101.0932}{{\tt
  arXiv:1101.0932}}].

\bibitem{Waxman:1996}
E.~{Waxman} and J.~{Miralda-Escude}, {\it {Images of Bursting Sources of
  High-Energy Cosmic Rays: Effects of Magnetic Fields}},  {\em Ap. J. Lett.} {\bf 472}
  (Dec., 1996) L89, [\href{http://arxiv.org/abs/arXiv:astro-ph/9607059}{{\tt
  arXiv:astro-ph/9607059}}].

\bibitem{yuksel+CenA12}
H.~{Yuksel}, T.~{Stanev}, M.~D. {Kistler}, and P.~P. {Kronberg}, {\it {The
  Centaurus A Ultrahigh-Energy Cosmic Ray Excess and the Local Extragalactic
  Magnetic Field}},  {\em ArXiv e-prints} (Mar., 2012)
  [\href{http://arxiv.org/abs/1203.3197}{{\tt arXiv:1203.3197}}].

\bibitem{Auger_energy:2009}
J.~Abraham {\em et~al.}, {\it {The Cosmic Ray Energy Spectrum and Related
  Measurements with the Pierre Auger Observatory}},  in {\em 31st ICRC}, June,
  2008.
\end{thebibliography}
\end{document}